\documentclass[letterpaper, english, aps, prl, reprint, twocolumn, superscriptaddress, floatfix]{revtex4-2}
\usepackage[T1]{fontenc}
\usepackage{color}
\usepackage{amsmath}
\usepackage{amssymb}
\usepackage{bm}
\usepackage{graphicx}
\usepackage{multirow}
\usepackage{booktabs}
\usepackage{capt-of}
\usepackage{babel}
\usepackage{calc}
\usepackage{dcolumn}
\usepackage{gensymb}
\usepackage{etoolbox}
\makeatletter
\usepackage[colorlinks=true, letterpaper=true, pdfstartview=FitV, linkcolor=blue, citecolor=blue, urlcolor=blue]{hyperref}
\renewcommand{\fnum@figure}{FIG. \thefigure}
\renewcommand{\fnum@table}{TABLE \thetable}
\makeatother

\newcommand{\bk}{\boldsymbol{k}}

\begin{document}
%========================================================
\title{Complete Hierarchy of Nonrelativistic Odd-Parity Spin Splitting in Collinear Magnets}
%\title{Collinear odd-parity magnets with all possible non-relativistic spin splitting ($\ell=1,3,5,7,9$)}
\author{Yichen Liu}
\author{Junxi Yu}
\author{Pu Zhang}
\author{Cheng-Cheng Liu}
\email{ccliu@bit.edu.cn}
\affiliation{Centre for Quantum Physics, Key Laboratory of Advanced Optoelectronic Quantum Architecture and Measurement (MOE), School of Physics, Beijing Institute of Technology, Beijing 100081, China}
%\affiliation{ address 3}
%\date{\today}
%=======================Abstract===============================================================
\begin{abstract}
Momentum-dependent nonrelativistic spin splitting provides a symmetry fingerprint of collinear magnets and can govern unconventional electronic, magnonic, and transport phenomena. Whereas even-parity $s$-, $d$-, $g$-, and $i$-wave splittings in collinear magnets have been extensively studied, odd-parity counterparts remain unexplored beyond the $p$-wave and $f$-wave classes. Here, using group theory, we establish the complete classification of odd-parity spin splitting in collinear magnets. We show that, in addition to the $p$-wave and $f$-wave forms, $h$- and $k$-wave splittings with $\ell=5$ and $7$ are allowed, while $m$-wave splitting with $\ell=9$ constitutes the upper bound. We derive a complete mapping from crystallographic point-group irreducible representations to the lowest-order odd-parity basis functions and formulate the coupling rule between a symmetry-breaking axial field and the parent N\'eel order that selects the induced odd-parity class. We further construct minimal lattice models that realize $h$-, $k$-, and $m$-wave splitting. Guided by this classification, we screen the MAGNDATA database and show that circularly polarized light can drive the $\mathcal{PT}$-symmetric antiferromagnets Fe$_2$TeO$_6$ and MgFe$_6$Ge$_6$ into $h$-wave and $k$-wave phases, respectively, exhibiting the hallmark spin splittings in both electronic bands and magnon spectra. Symmetry analysis and Berry-curvature calculations show that collinear odd-parity magnets of both $h$- and $k$-wave allow an anomalous Hall response, whereas the $m$-wave class forbids it. Together, these results complete the partial-wave hierarchy of odd-parity spin splitting in collinear magnets and establish symmetry criteria for anomalous transport in the high-partial-wave classes.
\end{abstract}

\maketitle

\textit{Introduction.---}
Momentum-dependent nonrelativistic spin splitting is redefining the landscape of collinear magnetism, with its parity, nodal structure, and partial-wave order providing symmetry fingerprints of electronic bands and bosonic excitations~\cite{yamadaMetallicPwaveMagnet2025,hellenesPwaveMagnets2023,linOddparityAltermagnetismSublattice2026,ezawaThirdorderFifthorderNonlinear2025,brekkeMinimalModelsTransport2024,yuOddParityMagnetismDriven2025,songElectricalSwitchingPwave2025,liuSpinGroupSymmetryMagnetic2022,smejkalConventionalFerromagnetismAntiferromagnetism2022,smejkalEmergingResearchLandscape2022,wuFermiLiquidInstabilities2007a,hayamiMomentumDependentSpinSplitting2019b,yuanGiantMomentumdependentSpin2020b,maMultifunctionalAntiferromagneticMaterials2021,kawamuraCompensatedFerrimagnetsColossal2024,liuTwoDimensionalFullyCompensated2025,smejkalChiralMagnonsAltermagnetic2023,cuiEfficientSpinSeebeck2023,liuChiralSplitMagnon2024a,chenUnconventionalMagnonsCollinear2025,wuMagnonSplittingTransport2025,jinInteractionDrivenAltermagnetic2026,searsAltermagneticDipolarSplitting2026,liuObservationSwitchableChiral2026,xieGeneralTheoryChiral2026a,zhangOddParityMagnons2026,luoSpinGroupSymmetry2026,yangSymmetryguidedCatalogueChiral2026,bendinDWavePhononAngular2026,wangAlteraxialPhononsCollinear2026}. Collinear magnets with even-parity spin splitting, including altermagnets and fully compensated ferrimagnets, have been intensively investigated~\cite{smejkalConventionalFerromagnetismAntiferromagnetism2022,smejkalEmergingResearchLandscape2022,wuFermiLiquidInstabilities2007a,hayamiMomentumDependentSpinSplitting2019b,yuanGiantMomentumdependentSpin2020b,maMultifunctionalAntiferromagneticMaterials2021,kawamuraCompensatedFerrimagnetsColossal2024,liuTwoDimensionalFullyCompensated2025}. By contrast, their odd-parity counterparts remain comparatively unexplored, and a symmetry-complete classification has yet to be established. This missing classification leaves unresolved the full hierarchy of odd-parity spin-splitting states, the symmetry conditions required to realize them, and the physical responses associated with different partial-wave classes.

\begin{figure}[t]
\centering
\includegraphics[width=0.48\textwidth]{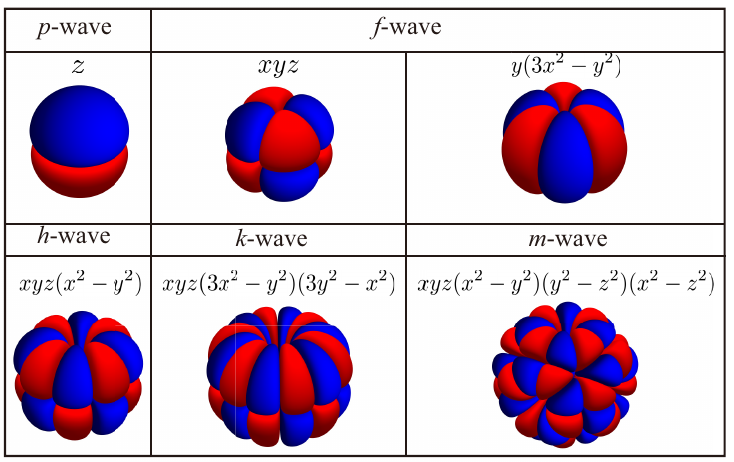}
\caption{Complete hierarchy of nonrelativistic odd-parity spin splitting in collinear magnets. The surfaces visualize the angular dependence of the partial-wave form factor of $\Delta E(\bk)=E^{\uparrow}(\bk)-E^{\downarrow}(\bk)$, with red and blue denoting positive and negative signs. The upper row shows the $p$-wave basis $z$ and representative $f$-wave bases $xyz$ and $y(3x^2-y^2)$; the lower row shows the $h$-, $k$-, and $m$-wave bases $xyz(x^2-y^2)$, $xyz(3x^2-y^2)(3y^2-x^2)$, and $xyz(x^2-y^2)(y^2-z^2)(x^2-z^2)$.}
\label{fig:hierarchy}
\end{figure}

In collinear magnets without spin-orbit coupling, $S_z$ is conserved, allowing the electronic structure to be resolved into two independent spin sectors, $s=\uparrow,\downarrow$. The spin-preserving effective time-reversal symmetry $[C_2\mathcal T\Vert E]$ enforces band energy $E^s(\bk)=E^s(-\bk)$, therefore restricts the spin splitting to be even in momentum. In parity-time ($\mathcal{PT}$)-symmetric antiferromagnets (AFMs), the spin-flipping inversion $[C_2\Vert P]$ additionally imposes $E^\uparrow(\bk)=E^\downarrow(-\bk)$. Together, these two symmetries guarantee spin degeneracy at every $\bk$. Breaking $[C_2\mathcal T\Vert E]$ while preserving $[C_2\Vert P]$ lifts the degeneracy but necessarily gives $\Delta E(\bk)=-\Delta E(-\bk)$, thereby generating odd-parity spin splitting~\cite{huangLightInducedOddParityMagnetism2026,liFloquetSpinSplitting2026,zhuFloquetOddParityCollinear2026,liuLightinducedOddparityAltermagnets2026}. Existing studies of collinear odd-parity magnets have focused primarily on $p$- and $f$-wave splitting. It remains unknown whether higher odd partial waves can occur in collinear magnets, whether their hierarchy possesses an upper bound, and how the representation of a parent Néel order combines with that of an external symmetry-breaking axial field to select a particular odd-parity form factor. Moreover, because breaking effective time-reversal symmetry permits time-reversal-odd transport, it is important to determine which high-partial-wave classes can support an anomalous Hall response.

In this Letter, we establish the complete hierarchy of nonrelativistic odd-parity spin splitting in collinear magnets. As plotted in Fig.~\ref{fig:hierarchy}, we show that all allowed classes are  $p$, $f$, $h$, $k$ and $m$-wave, corresponding to partial-wave orders $\ell=1,3,5,7,9$, respectively, with $\ell=9$ constituting the upper bound. We provide a complete mapping between the spin-splitting irreducible representations and their lowest-order odd-parity basis functions, and derive the coupling rule that relates the induced spin-splitting representation to the product of the parent N\'{e}el-order representation and the axial-vector representation of an external symmetry-breaking field. We further construct minimal tight-binding models realizing the previously unexplored $h$-, $k$- and $m$-wave classes. Guided by this classification, we screen the MAGNDATA database for candidate materials and find that circularly-polarized-light irradiated Fe$_2$TeO$_6$ and MgFe$_6$Ge$_6$ can be odd-parity magnets of $h$-wave and $k$-wave, respectively, with the characteristic splittings in their electronic band structure and magnon spectra. Among the high-partial-wave odd-parity magnets, $h$-wave and $k$-wave odd-parity magnets can host anomalous transport, whereas the high symmetry of the $m$-wave forbids it.

\textit{{Complete classification of the odd-parity spin splitting}.---}
We first classify odd-parity spin splitting in collinear magnets independently of its microscopic origin. Under a crystallographic point-group operation $\mathcal O$, the spin splitting transforms as 
\begin{equation}
\Delta E(\mathcal O\bk)=\chi^\Delta(\mathcal O)\Delta E(\bk),
\label{eq:classification}
\end{equation}
where $\chi^\Delta(\mathcal O)$ is the character of the one-dimensional irreducible representation (irrep) $\Gamma_\Delta$ carried by $\Delta E(\bk)$. Its partial-wave class is determined by the lowest-order nonvanishing odd polynomial in the expansion around the $\Gamma$ point, subject to $\Delta E(-\bk)=-\Delta E(\bk)$.  

For a collinear spin point group, the N\'{e}el order transforms according to a one-dimensional irrep $\Gamma_N$ of the corresponding crystallographic point group~\cite{mcclartyLandauTheoryAltermagnetism2024,schiffCollinearAltermagnetsTheir2025,liuAntiferroaxialAltermagnetism2026a}. This order-parameter representation must be distinguished from the spin-splitting representation $\Gamma_\Delta$. A controllable route to breaking the effective time-reversal symmetry $[C_2\mathcal{T}\Vert E]$ is Floquet driving by circularly polarized light~\cite{goldmanPeriodicallyDrivenQuantum2014,okaFloquetEngineeringQuantum2019,huangLightInducedOddParityMagnetism2026,liFloquetSpinSplitting2026,zhuFloquetOddParityCollinear2026,liuLightinducedOddparityAltermagnets2026}. We write the vector potential as $\boldsymbol{\mathcal A}(t)=\operatorname{Re}[\boldsymbol{\mathcal A}_0e^{-i\omega t}]$, $\boldsymbol{\mathcal A}_0=\mathcal A_0(\hat{\mathbf x}+i\eta\hat{\mathbf y})$, where $\mathcal A_0$ is the field amplitude and $\eta=\pm1$ specifies the helicity. The handedness of the drive is represented by the time-reversal-odd axial vector
\begin{equation}
\boldsymbol\kappa=i\boldsymbol{\mathcal A}_0\times\boldsymbol{\mathcal A}_0^*,
\label{eq:Floquetdriving}
\end{equation}
which transforms according to the axial-vector representation $\Gamma_A$. The symmetry channels available to the induced spin splitting therefore satisfy
\begin{equation}
\Gamma_\Delta\subseteq\Gamma_A\otimes\Gamma_N.
\label{eq:classification2}
\end{equation}
(see details in Supplemental Material (SM)~\cite{SM}). The parent collinear spin point group fixes $\Gamma_N$, while the propagation direction selects the relevant component of $\Gamma_A$. Their direct product then determines the allowed $\Gamma_\Delta$, which in turn fixes the lowest-order form factor and its partial-wave order $\ell$.

Table~\ref{table:odd_parity} lists the lowest-order odd-parity basis functions for every one-dimensional spin-splitting irrep $\Gamma_\Delta$ of the crystallographic point groups. For each relevant parent collinear spin point group, Table~\ref{table:floquet_coupling} gives the N\'{e}el-order irrep $\Gamma_N$, the decomposition of the axial-vector representation $\Gamma_A$, and the resulting product $\Gamma_N\otimes\Gamma_A$. The decomposition of $\Gamma_A$ distinguishes symmetry-inequivalent orientations of the axial field, including the principal-axis and transverse components. Combining the two tables provides an exhaustive two-step classification: Table~\ref{table:floquet_coupling} determines the allowed spin-splitting representation for a given parent magnetic order and field direction, whereas Table~\ref{table:odd_parity} maps that representation onto its lowest-order odd-parity form factor. In addition to the previously known $p$- and $f$-wave classes with $\ell=1$ and $3$, respectively, this procedure yields new $h$-, $k$-, and $m$-wave classes with $\ell=5, 7,$ and $9$. Across all crystallographic point groups, no one-dimensional odd-parity representation first appears above degree nine, establishing $\ell=9$ as the upper bound.

\textit{Minimal symmetry-adapted tight-binding models.---}
To verify the classification and provide an explicit microscopic foundation, we build minimal four-band Hamiltonians with two magnetic sublattices and two spin channels,
\begin{equation}
H(\bk)=d_0(\bk)+\bm d(\bk)\cdot\bm\tau+J\tau_z\sigma_z,
\label{eq:model}
\end{equation}
where $\bm\tau$ and $\bm\sigma$ act on the sublattice and spin degrees of freedom, and $J\tau_z\sigma_z$ is the staggered exchange field associated with the N\'eel order. The lattice geometries, sublattice positions, and real-space hopping patterns of the minimal models are detailed in SM~\cite{SM}. For a spin eigenvalue $s=\pm1$, the band energies are
\begin{equation}
E_{\nu s}(\bk)=d_0(\bk)+\nu\sqrt{d_x^2(\bk)+d_y^2(\bk)+[d_z(\bk)+sJ]^2},
\label{eq:model_spectrum}
\end{equation}
where $\nu=\pm1$ is the band index. From Eq.~\eqref{eq:model_spectrum}, the exact spin splitting is $\Delta E_\nu(\bk)=4\nu Jd_z(\bk)/\mathcal N(\bk)$, where $\mathcal N(\bk)$ is even in $d_z$ and invariant under all point-group operations (see details in SM~\cite{SM}). Hence $\Delta E_\nu(\bk)$ and the sublattice-contrasting hopping $d_z(\bk)$ belong to the same irrep.

This result clarifies the role of each term in the four-band construction: $d_x$ and $d_y$ hybridize the two magnetic sublattices and produce the basic band dispersion, $J$ generates opposite exchange fields on the two sublattices, and $d_z(\bk)$ provides the momentum-dependent sublattice contrast with the desired symmetry. Neither $J$ nor $d_z$ alone produces the target spin splitting: the former is momentum independent, whereas the latter is spin independent. Their coexistence converts the form factor carried by $d_z$ into a momentum-dependent spin splitting. 
Constructing a given odd-parity class therefore reduces to identifying a spin-independent sublattice-contrasting hopping $d_z(\bk)$ that transforms according to the desired $\Gamma_\Delta$.

\begin{figure}[t]
\centering
\includegraphics[width=0.48\textwidth]{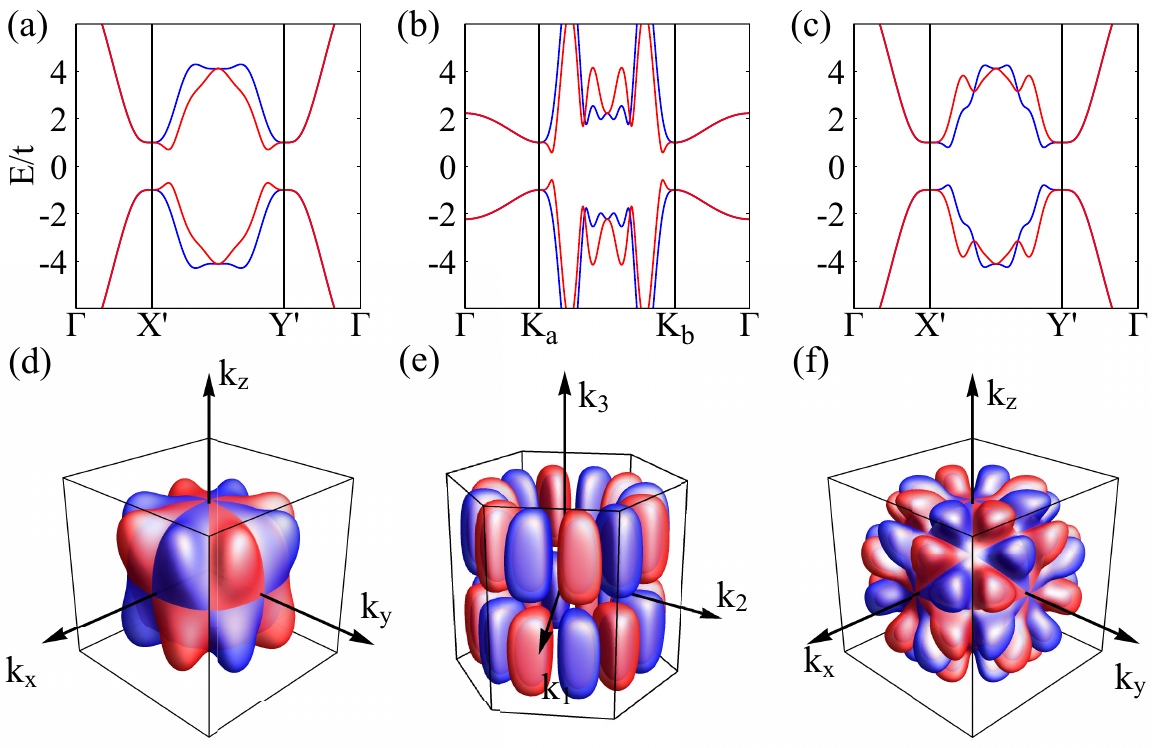}
\caption{Minimal symmetry-adapted models for high-partial-wave odd-parity spin splitting. Panels (a)-(c) show the spin-resolved band structures of the $h$-, $k$-, and $m$-wave models, respectively, with red and blue denoting the spin-up and spin-down sectors. Panels (d)-(f) show the corresponding isoenergy surfaces at $E=-3t$ in the cubic, hexagonal, and cubic Brillouin zones, respectively; the colors distinguish the two spin channels. In each model, the nearest-neighbor hopping amplitude $t$ is taken as the energy unit. The parameters are $J=t$, $\lambda_h=0.2t$, $\lambda_k=0.5t$, and $\lambda_m=0.1t$.}
\label{fig:tb_model}
\end{figure}

For representative $p$- and $f$-wave cases, together with the cubic $h$-wave, hexagonal $k$-wave, and cubic $m$-wave cases
\begin{align}
d_z^p(\bk)&=2\lambda\sin k_z,\notag\\
d_z^{f_b}(\bk)&=8\lambda\sin k_x\sin k_y\sin k_z,\notag\\
d_z^{f_p}(\bk)&=8\lambda \sin k_1\sin k_2\sin(k_1+k_2),\notag\\
d_z^h(\bk)&=16\lambda_h(\cos k_x-\cos k_y)\prod_{i=x,y,z}\sin k_i,
\notag\\
d_z^k(\bk)&=64\lambda_k\sin k_3\sin \frac{k_1}{2}\sin \frac{k_2}{2}(\cos k_2-\cos k_1)
\notag\\
&\quad\times\sin\left(k_1+\frac{k_2}{2}\right)\sin\left(\frac{k_1}{2}+k_2\right),
\notag\\
d_z^m(\bk)&=64\lambda_m\prod_{\mathrm cyc}\sin k_i(\cos k_j-\cos k_l).
\label{eq:model_form_factors}
\end{align}
Here, $k_1$, $k_2$, and $k_3$ are the symmetry-adapted momentum coordinates of the hexagonal model, and the final product runs over the cyclic permutations $(i,j,l)=(x,y,z),(y,z,x),(z,x,y)$ (see details in SM~\cite{SM}). The labels $f_b$ and $f_p$ denote the $xyz$ and $x(x^2-3y^2)$ forms of $f$-wave splitting, respectively. Expanding Eq.~\eqref{eq:model_form_factors} around the $\Gamma$ point gives $d_z^p\propto k_z$, $d_z^{f_b}\propto k_xk_yk_z$, $d_z^{f_p}\propto k_x(k_x^2-3k_y^2)$, $d_z^h\propto k_xk_yk_z(k_x^2-k_y^2)$, $d_z^k\propto k_xk_yk_z(3k_x^2-k_y^2)(k_x^2-3k_y^2)$, and $d_z^m\propto k_xk_yk_z(k_x^2-k_y^2)(k_y^2-k_z^2)(k_x^2-k_z^2)$. These long-wavelength limits reproduce the basis functions listed in Table~\ref{table:odd_parity} and confirm the corresponding $p$-, $f$-, $h$-, $k$-, and $m$-wave characters. The corresponding spin-resolved band structures and isoenergy surfaces for the $h$-, $k$-, and $m$-wave models are shown in Fig.~\ref{fig:tb_model}.

The imaginary hopping channels in these models break the effective time-reversal symmetry and may arise from an applied magnetic flux, loop-current order, or periodic driving. Although the models above are constructed for electronic bands, the same representation-based procedure applies to even-parity spin splitting and can also be extended to odd-parity magnon bands. Corresponding minimal four-band bosonic Hamiltonians for high-partial-wave odd-parity magnons are provided in SM~\cite{SM}.

The six form-factor terms in Eq.~\eqref{eq:model_form_factors} are written in their maximal-symmetry limits. Because the imaginary $d_z$ channels are odd under the effective time-reversal operation, the fixed-spin magnetic point groups are $4/m'mm$, $4'/m'mm'$, $\bar{6}m'2'$, $4/m'm'm'$, $6/m'm'm'$, and $m'\bar{3}'m'$ for the $p$-, $f_b$-, $f_p$-, $h$-, $k$-, and $m$-wave forms, respectively (see details in SM~\cite{SM}). In a real material, however, the symmetry of a fixed spin sector is obtained by combining the spin-preserving $[E\Vert \mathcal O]$ subgroup of the collinear spin point group with the time-reversal-breaking external perturbation, such as circularly polarized Floquet driving. Additional spin-independent hoppings entering $d_x$ and $d_y$ can therefore lower the idealized models, when necessary, to the material-compatible fixed-spin magnetic point groups $42'2'$, $\bar{4}2'm'$, $\bar{6}m'2'$, $4m'm'$, $6m'm'$, and $4'32'$ for the $p$-, $f_b$-, $f_p$-, $h$-, $k$-, and $m$-wave forms, respectively.
Crucially, these additional hoppings do not alter the symmetry or momentum dependence of the spin splitting.

A simpler phenomenological construction would start from the continuum basis functions in Table~\ref{table:odd_parity} and directly regularize them on a lattice through the substitutions $k_i\rightarrow\sin k_i$ and $k_i^2\rightarrow 2(1-\cos k_i)$. Although this procedure preserves the low-energy expansion around the $\Gamma$ point, it generally produces artificial, explicitly spin-dependent hopping terms. In contrast, our construction keeps all nontrivial hopping terms spin independent: the spin splitting emerges only through their coupling to the staggered exchange field. It therefore provides a more transparent microscopic realization of the symmetry-classified odd-parity form factors.

\textit{{Material candidates}.---}
Tables~\ref{table:odd_parity} and \ref{table:floquet_coupling} provide a direct symmetry-based route for identifying material realizations: the parent collinear spin point group determines the N\'{e}el-order representation $\Gamma_N$, while the direction of the applied axial field selects the relevant component of $\Gamma_A$; their product then fixes the induced spin-splitting representation $\Gamma_\Delta$ and its lowest-order form factor. Guided by this framework, we screen the MAGNDATA database~\cite{gallegoMAGNDATADatabaseMagnetic2016} and identify tetragonal Fe$_2$TeO$_6$ and hexagonal MgFe$_6$Ge$_6$ as representative candidates. Both are intrinsic $\mathcal{PT}$-symmetric collinear AFMs and are spin degenerate in equilibrium.

\begin{figure*}[t]
\centering
\includegraphics[width=0.98\textwidth]{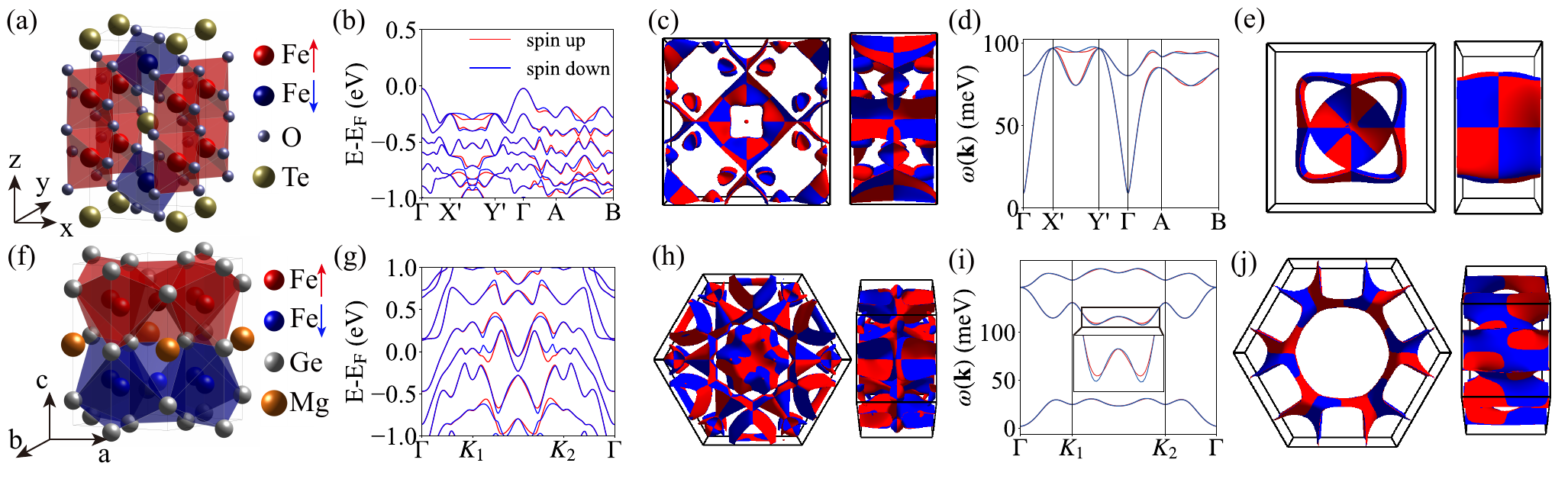}
\caption{Material realizations of high-partial-wave collinear odd-parity magnetism. (a) Tetragonal crystal structure [space group $P4_2/mnm$ (No.~136)] and $\mathcal{PT}$-symmetric collinear antiferromagnetic (AFM) order of Fe$_2$TeO$_6$; red and blue Fe atoms carry opposite magnetic moments. (b) Spin-resolved electronic band structure and (c) two views of the corresponding electronic isoenergy surfaces in the light-induced $h$-wave phase. (d) Magnon spectrum and (e) two views of the corresponding spin-resolved magnon isoenergy surfaces. (f) Hexagonal crystal structure [space group $P6/mmm$ (No.~191)] and $\mathcal{PT}$-symmetric collinear AFM order of MgFe$_6$Ge$_6$. Panels (g) and (h) show the spin-resolved electronic band structure and isoenergy surfaces, respectively, in the light-induced $k$-wave phase; panels (i) and (j) show the corresponding magnon spectrum and spin-resolved magnon isoenergy surfaces. The inset in panel (i) magnifies the magnon splitting. Red and blue denote the two spin sectors. The circularly polarized light propagates along $z$, with $e\mathcal A_0/\hbar=0.2~\text{\AA}^{-1}$ and $\hbar\omega=1~\mathrm{eV}$.}
\label{fig:materials_realization}
\end{figure*}

To break the effective time-reversal symmetry $[C_2\mathcal T\Vert E]$, we apply circularly polarized light propagating along the crystal $z$ axis. Taking $\eta=+1$, the vector potential is $\boldsymbol{\mathcal A}(t)=\mathcal A_0(\cos\omega t,\sin\omega t,0)$. The corresponding component $\kappa_z$ transforms as the axial representation $A_{2g}$ in both $4/mmm$ and $6/mmm$. The drive breaks $[C_2\mathcal T\Vert E]$ while preserving the spin-flipping inversion $[C_2\Vert P]$. The former lifts the spin degeneracy, while the latter retains the relation $E^\uparrow(\bk)=E^\downarrow(-\bk)$ and hence enforces $\Delta E(\bk)=-\Delta E(-\bk)$~\cite{zhuFloquetOddParityCollinear2026,eckardtColloquiumAtomicQuantum2017}. Details of the Floquet-Wannier construction and numerical settings are provided in SM~\cite{SM}.

The experimentally synthesized Fe$_2$TeO$_6$ has the crystal structure and collinear magnetic order shown in Fig.~\ref{fig:materials_realization}(a)~\cite{kunnmannMagneticStructuresOrdered1968}, where red and blue Fe atoms denote opposite magnetic moments. Its parent spin point group is $4/{}^{\bar{1}}mmm$, corresponding to $\Gamma_N=A_{2u}$ (Table~\ref{table:floquet_coupling}). For light propagating along $z$, $\Gamma_A=A_{2g}$, giving $\Gamma_A\otimes\Gamma_N=A_{2g}\otimes A_{2u}=A_{1u}$, whose lowest-order odd basis in the crystallographic point group $4/mmm$ is the $h$-wave form $k_xk_yk_z(k_x^2-k_y^2)$ (Table~\ref{table:odd_parity}). 
Within a fixed spin sector, the spin-preserving spatial symmetry of the equilibrium state is $4mm$. The Floquet component $\kappa_z$ transforms as the time-reversal-odd axial order parameter $A_2^{-}$ and lowers the fixed-spin symmetry to the magnetic point group $4m'm'$. As shown in Fig.~\ref{fig:materials_realization}(b), the circular drive lifts the equilibrium spin degeneracy, while the isoenergy surfaces in Fig.~\ref{fig:materials_realization}(c) display the characteristic sign reversals and nodal planes of the $h$-wave splitting.

The experimentally synthesized MgFe$_6$Ge$_6$, whose crystal and collinear magnetic structures are shown in Fig.~\ref{fig:materials_realization}(f)~\cite{mazetMagneticPropertiesMgFe6Ge62013}, has the parent spin point group $6/{}^{\bar{1}}mmm$ and likewise transforms as $\Gamma_N=A_{2u}$. For a $z$-directed axial field, $\Gamma_A=A_{2g}$, so that $\Gamma_A\otimes\Gamma_N=A_{2g}\otimes A_{2u}=A_{1u}$.
In $6/mmm$, however, the lowest-order odd basis of $A_{1u}$ is $k_xk_yk_z(3k_x^2-k_y^2)(k_x^2-3k_y^2)$, corresponding to $k$-wave splitting.
The spin-preserving symmetry within each fixed spin sector is $6mm$, and the $A_2^{-}$ Floquet helicity lowers it to $6m'm'$. The resulting electronic band structure in Fig.~\ref{fig:materials_realization}(g) exhibits a finite nonrelativistic spin splitting, while the isoenergy surfaces in Fig. ~\ref{fig:materials_realization}(h) reveal its characteristic sixfold in-plane angular pattern.

The same symmetry mechanism also applies to bosonic excitations. In a collinear AFM, the two magnon branches with opposite chirality play roles analogous to the two electronic spin sectors. Circular driving breaks the corresponding effective time-reversal symmetry and induces magnon splitting with the same spatial representation and partial-wave form factor as in the electronic bands. Figures~\ref{fig:materials_realization}(d) and \ref{fig:materials_realization}(e) show the resulting $h$-wave magnon splitting and spin-resolved isoenergy surfaces in Fe$_2$TeO$_6$, whereas Figs.~\ref{fig:materials_realization}(i) and \ref{fig:materials_realization}(j) display the corresponding $k$-wave magnon spectra and isoenergy surfaces in MgFe$_6$Ge$_6$.

\textit{{Anomalous transport}.---}
Breaking time-reversal symmetry within a fixed spin sector permits time-reversal-odd responses, most notably the anomalous Hall effect. Whether such a response is allowed is determined not by the partial-wave order alone, but by the magnetic point group of the driven state. For Fe$_2$TeO$_6$ and MgFe$_6$Ge$_6$, the fixed-spin magnetic point groups $4m'm'$ and $6m'm'$, respectively, allow an axial Hall vector along $z$ and hence a finite anomalous Hall conductivity in the $xy$ plane. By contrast, the cubic $m$-wave class has fixed-spin magnetic point group $4'32'$, which forbids every component of the anomalous Hall conductivity (see details in SM~\cite{SM}). Thus, among the five symmetry-allowed odd-parity classes, the $p$-, $f$-, $h$-, and $k$-wave classes can support anomalous Hall response, whereas the $m$-wave class cannot.

\begin{figure}[t]
\centering
\includegraphics[width=0.48\textwidth]{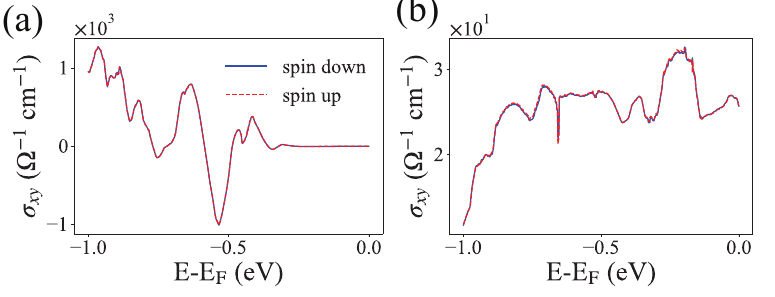}
\caption{Anomalous Hall conductivity of high-partial-wave odd-parity magnets. Spin-resolved anomalous Hall conductivity as a function of the chemical-potential shift $E-E_{\mathrm F}$ for (a) $h$-wave Fe$_2$TeO$_6$ and (b) $k$-wave MgFe$_6$Ge$_6$. Blue solid and red dashed curves denote the spin-down and spin-up contributions, respectively. The light-field parameters are the same as in Fig.~\ref{fig:materials_realization}.}
\label{fig:transport}
\end{figure}

For spin sector $s$, the intrinsic anomalous Hall conductivity is
$\sigma_{xy}^{s}=
-\frac{e^2}{\hbar}
\sum_n
\int_{\mathrm{BZ}}
\frac{d^3k}{(2\pi)^3}
f_{ns}(\bk)\Omega^z_{ns}(\bk)$~\cite{yaoFirstPrinciplesCalculation2004},
where $f_{ns}(\bk)$ is the Fermi--Dirac distribution and $\Omega_{ns}^z(\bk)$ is the Berry curvature of the $n$th band in spin sector $s$. Once the effective time-reversal symmetry $[C_2\mathcal T\Vert E]$ is broken by the circular drive, the total charge Hall response is the sum of the two spin-sector contributions. Figure~\ref{fig:transport}(a) shows the calculated Hall conductivity of the $h$-wave Fe$_2$TeO$_6$. A pronounced peak appears in the valence-band region, where the contribution from each spin sector reaches approximately $1.5\times10^3~\Omega^{-1}\mathrm{cm}^{-1}$. This large response originates from the enhanced Berry curvature near the nodal-line band structure. In metallic MgFe$_6$Ge$_6$, Fig.~\ref{fig:transport}(b) shows a finite Hall conductivity near the Fermi energy, demonstrating that the $k$-wave phase can likewise produce an experimentally accessible charge Hall response. These results show that high-partial-wave odd-parity spin splitting is compatible with anomalous transport whenever permitted by the residual magnetic symmetry.

\textit{Conclusion and discussion.---}
We have established a complete symmetry classification of nonrelativistic odd-parity spin splitting in collinear magnets. Beyond the known $p$-wave and $f$-wave classes, we identify the $h$, $k$, and $m$-wave classes and show that the hierarchy terminates at partial-wave order $\ell=9$. 
Tables~\ref{table:odd_parity} and \ref{table:floquet_coupling} provide a direct two-step procedure for determining the induced spin splitting: for a given parent collinear spin point group and axial-field direction, Table~\ref{table:floquet_coupling} determines the allowed components of $\Gamma_A\otimes\Gamma_N$, while Table~\ref{table:odd_parity} maps each resulting one-dimensional spin-splitting representation $\Gamma_\Delta$ onto its lowest-order odd-parity basis function and partial-wave order. We further constructed minimal symmetry-adapted four-band models realizing all five odd-parity classes. In these models, the spin splitting emerges from the interplay between a spin-independent, momentum-dependent sublattice contrast and the staggered exchange field. Applying the classification to the MAGNDATA database, we identify irradiated Fe$_2$TeO$_6$ and MgFe$_6$Ge$_6$ as representative $h$- and $k$-wave odd-parity magnets with the characteristic high-partial-wave structures appearing in both their electronic and magnon spectra. Moreover, the residual magnetic symmetries permit anomalous Hall responses in the $h$- and $k$-wave phases, whereas the cubic $m$-wave symmetry forbids all components of the anomalous Hall conductivity. This classification and mechanism can be directly applied to other bosonic excitations carrying suitable angular-momentum or pseudospin degrees of freedom in collinear odd-parity magnets, especially the high-partial-wave ones.

Although we have focused on circularly polarized light as a controllable means of breaking the effective time-reversal symmetry, the classification is independent of the microscopic symmetry-breaking mechanism and also applies to magnetic flux, loop-current order, and other time-reversal-odd axial perturbations. The driving parameters considered here correspond to short, intense pump pulses with a peak electric field of approximately $0.2~\mathrm{V\,\text{\AA}^{-1}}$; experimental realizations will require control of the off-resonant condition, heating, and pulse duration~\cite{eckardtColloquiumAtomicQuantum2017,deLaTorreColloquiumNonthermal2021}. We have not identified a solid-state realization of the cubic $m$-wave class in the current version of the MAGNDATA database. Cold atoms provide a promising alternative, because lattice geometry, tunneling amplitudes, interactions, and Peierls phases can be controlled independently, while laser-assisted tunneling and periodic modulation enable the engineering of complex long-range hoppings~\cite{blochManybodyPhysicsUltracold2008,miyakeRealizingHarperHamiltonian2013,jotzuExperimentalRealizationTopological2014,eckardtColloquiumAtomicQuantum2017,dasRealizingAltermagnetismFermiHubbard2024}. Such platforms may therefore offer a direct route to realizing the maximal $\ell=9$ odd-parity spin-splitting class.

{
\textit{Note added.---}
During completion of this manuscript, we became aware of~\cite{elcoroSpinPointGroup2026}, which classifies the nonrelativistic spin splitting in noncollinear magnets and differs from our manuscript, which focuses on collinear magnets with the effective time-reversal symmetry broken.
}

\bibliography{ref}

@article{liuSpinGroupSymmetryMagnetic2022,
  title = {Spin-{{Group Symmetry}} in {{Magnetic Materials}} with {{Negligible Spin-Orbit Coupling}}},
  author = {Liu, Pengfei and Li, Jiayu and Han, Jingzhi and Wan, Xiangang and Liu, Qihang},
  date = {2022-04-21},
  year = 2022,
  journal = {Physical Review X},
  shortjournal = {Phys. Rev. X},
  volume = {12},
  number = {2},
  pages = {021016},
  doi = {10.1103/PhysRevX.12.021016}
}

@article{smejkalEmergingResearchLandscape2022,
  title = {Emerging {{Research Landscape}} of {{Altermagnetism}}},
  author = {{\v S}mejkal, Libor and Sinova, Jairo and Jungwirth, Tomas},
  date = {2022-12-08},
  year = 2022,
  journal = {Physical Review X},
  shortjournal = {Phys. Rev. X},
  volume = {12},
  number = {4},
  pages = {040501},
  doi = {10.1103/PhysRevX.12.040501}
}

@misc{luoSpinGroupSymmetry2026,
  title = {Spin {{Group Symmetry Criteria}} for {{Odd-parity Magnets}}},
  author = {Luo, Xun-Jiang and Hu, Jin-Xin and Hu, Meng-Li and Law, K. T.},
  year = 2026,
  number = {arXiv:2510.05512},
  eprint = {2510.05512},
  publisher = {arXiv},
  doi = {10.48550/arXiv.2510.05512},
  archiveprefix = {arXiv}
}

@article{kawamuraCompensatedFerrimagnetsColossal2024,
  title = {Compensated Ferrimagnets with Colossal Spin Splitting in Organic Compounds},
  author = {Kawamura, Taiki and Yoshimi, Kazuyoshi and Hashimoto, Kenichiro and Kobayashi, Akito and Misawa, Takahiro},
  year = 2024,
  month = apr,
  journal = {Physical Review Letters},
  volume = {132},
  number = {15},
  pages = {156502},
  issn = {0031-9007, 1079-7114},
  doi = {10.1103/PhysRevLett.132.156502},
  urldate = {2024-06-26},
  langid = {english}
}

@article{liuTwoDimensionalFullyCompensated2025,
  title = {Two-Dimensional Fully Compensated Ferrimagnetism},
  author = {Liu, Yichen and Guo, San-Dong and Li, Yongpan and Liu, Cheng-Cheng},
  year = 2025,
  month = mar,
  journal = {Physical Review Letters},
  volume = {134},
  number = {11},
  pages = {116703},
  publisher = {American Physical Society},
  doi = {10.1103/PhysRevLett.134.116703},
  urldate = {2025-06-25}
}

@article{yangSymmetryguidedCatalogueChiral2026,
  title = {Symmetry-Guided Catalogue of Chiral Phonon Materials},
  author = {Yang, Yue and Xiao, Zhenyu and Mao, Yu and Li, Zhanghuan and Wang, Zhenyang and Deng, Tianqi and Tang, Yanhao and Song, Zhi-Da and Li, Yuan and Yuan, Huiqiu and Shi, Ming and Xu, Yuanfeng},
  year = 2026,
  month = jun,
  journal = {Nature Physics},
  volume = {22},
  number = {6},
  pages = {884--890},
  publisher = {Nature Publishing Group},
  issn = {1745-2481},
  doi = {10.1038/s41567-026-03260-0},
  urldate = {2026-07-21},
  copyright = {2026 The Author(s), under exclusive licence to Springer Nature Limited},
  langid = {english}
}

@misc{bendinDWavePhononAngular2026,
  title = {D-{{Wave Phonon Angular Momentum Texture}} in {{Altermagnets}} by {{Magnon-Phonon-Hybridization}}},
  author = {Bendin, Hannah and Mook, Alexander and Mertig, Ingrid and Neumann, Robin R.},
  year = 2026,
  month = mar,
  number = {arXiv:2511.08357},
  eprint = {2511.08357},
  primaryclass = {cond-mat.mes-hall},
  publisher = {arXiv},
  doi = {10.48550/arXiv.2511.08357},
  urldate = {2026-07-21},
  archiveprefix = {arXiv}
}

@misc{wangAlteraxialPhononsCollinear2026,
  title = {Alteraxial {{Phonons}} in {{Collinear Magnets}}},
  author = {Wang, Fuyi and Xu, Junqi and Liu, Xinqi and Wang, Huaiqiang and Zhang, Lifa and Zhang, Haijun},
  year = 2026,
  month = jan,
  number = {arXiv:2512.07518},
  eprint = {2512.07518},
  primaryclass = {cond-mat.mtrl-sci},
  publisher = {arXiv},
  doi = {10.48550/arXiv.2512.07518},
  urldate = {2026-07-21},
  archiveprefix = {arXiv}
}

@article{smejkalConventionalFerromagnetismAntiferromagnetism2022,
  title = {Beyond Conventional Ferromagnetism and Antiferromagnetism: A Phase with Nonrelativistic Spin and Crystal Rotation Symmetry},
  shorttitle = {Beyond Conventional Ferromagnetism and Antiferromagnetism},
  author = {{\v S}mejkal, Libor and Sinova, Jairo and Jungwirth, Tomas},
  year = 2022,
  month = sep,
  journal = {Physical Review X},
  volume = {12},
  number = {3},
  pages = {031042},
  publisher = {American Physical Society},
  doi = {10.1103/PhysRevX.12.031042},
  urldate = {2023-03-15}
}

@article{wuFermiLiquidInstabilities2007a,
  title = {Fermi Liquid Instabilities in the Spin Channel},
  author = {Wu, Congjun and Sun, Kai and Fradkin, Eduardo and Zhang, Shou-Cheng},
  year = 2007,
  month = mar,
  journal = {Physical Review B},
  volume = {75},
  number = {11},
  pages = {115103},
  publisher = {American Physical Society},
  doi = {10.1103/PhysRevB.75.115103},
  urldate = {2026-01-29}
}

@article{hayamiMomentumDependentSpinSplitting2019b,
  title = {Momentum-Dependent Spin Splitting by Collinear Antiferromagnetic Ordering},
  author = {Hayami, Satoru and Yanagi, Yuki and Kusunose, Hiroaki},
  year = 2019,
  month = dec,
  journal = {Journal of the Physical Society of Japan},
  volume = {88},
  number = {12},
  pages = {123702},
  publisher = {The Physical Society of Japan},
  issn = {0031-9015},
  doi = {10.7566/JPSJ.88.123702},
  urldate = {2026-01-27}
}

@article{yuanGiantMomentumdependentSpin2020b,
  title = {Giant Momentum-Dependent Spin Splitting in Centrosymmetric Low-\$Z\$ Antiferromagnets},
  author = {Yuan, Lin-Ding and Wang, Zhi and Luo, Jun-Wei and Rashba, Emmanuel I. and Zunger, Alex},
  year = 2020,
  month = jul,
  journal = {Physical Review B},
  volume = {102},
  number = {1},
  pages = {014422},
  publisher = {American Physical Society},
  doi = {10.1103/PhysRevB.102.014422},
  urldate = {2026-01-27}
}

@article{maMultifunctionalAntiferromagneticMaterials2021,
  title = {Multifunctional Antiferromagnetic Materials with Giant Piezomagnetism and Noncollinear Spin Current},
  author = {Ma, Hai-Yang and Hu, Mengli and Li, Nana and Liu, Jianpeng and Yao, Wang and Jia, Jin-Feng and Liu, Junwei},
  year = 2021,
  month = may,
  journal = {Nature Communications},
  volume = {12},
  number = {1},
  pages = {2846},
  publisher = {Nature Publishing Group},
  issn = {2041-1723},
  doi = {10.1038/s41467-021-23127-7},
  urldate = {2024-03-02},
  copyright = {2021 The Author(s)},
  langid = {english}
}

@article{yamadaMetallicPwaveMagnet2025,
  title = {A Metallic {$p$}-Wave Magnet with Commensurate Spin Helix},
  author = {Yamada, Rinsuke and Birch, Max T. and Baral, Priya R. and Okumura, Shun and Nakano, Ryota and Gao, Shang and Ezawa, Motohiko and Nomoto, Takuya and Masell, Jan and Ishihara, Yuki and Kolincio, Kamil K. and Belopolski, Ilya and Sagayama, Hajime and Nakao, Hironori and Ohishi, Kazuki and Ohhara, Takashi and Kiyanagi, Ryoji and Nakajima, Taro and Tokura, Yoshinori and Arima, Taka-hisa and Motome, Yukitoshi and Hirschmann, Moritz M. and Hirschberger, Max},
  year = 2025,
  month = oct,
  journal = {Nature},
  volume = {646},
  number = {8086},
  pages = {837--842},
  publisher = {Nature Publishing Group},
  doi = {10.1038/s41586-025-09633-4}
}

@misc{hellenesPwaveMagnets2023,
  title = {{$p$}-Wave Magnets},
  author = {Hellenes, Anna Birk and Jungwirth, Tom{\'a}{\v s} and Jaeschke-Ubiergo, Rodrigo and Chakraborty, Atasi and Sinova, Jairo and {\v S}mejkal, Libor},
  year = 2023,
  publisher = {arXiv},
  doi = {10.48550/arXiv.2309.01607},
  eprint = {2309.01607},
  archiveprefix = {arXiv},
  primaryclass = {cond-mat.mes-hall}
}

@article{linOddparityAltermagnetismSublattice2026,
  title = {Odd-Parity Altermagnetism through Sublattice Currents: From {Haldane--Hubbard} Model to General Bipartite Lattices},
  author = {Lin, Yu-Ping and Vila, Marc},
  year = 2026,
  journal = {Physical Review Letters},
  publisher = {American Physical Society},
  doi = {10.1103/c8pd-2fs4},
  note = {Accepted 10 June 2026}
}

@article{ezawaThirdorderFifthorderNonlinear2025,
  title = {Third-Order and Fifth-Order Nonlinear Spin-Current Generation in {$g$}-Wave and {$i$}-Wave Altermagnets and Perfectly Nonreciprocal Spin Current in {$f$}-Wave Magnets},
  author = {Ezawa, Motohiko},
  year = 2025,
  month = mar,
  journal = {Physical Review B},
  volume = {111},
  number = {12},
  pages = {125420},
  publisher = {American Physical Society},
  doi = {10.1103/PhysRevB.111.125420}
}

@article{brekkeMinimalModelsTransport2024,
  title = {Minimal Models and Transport Properties of Unconventional {$p$}-Wave Magnets},
  author = {Brekke, Bj{\o}rnulf and Sukhachov, Pavlo and Giil, Hans Gl{\o}ckner and Brataas, Arne and Linder, Jacob},
  year = 2024,
  month = dec,
  journal = {Physical Review Letters},
  volume = {133},
  number = {23},
  pages = {236703},
  publisher = {American Physical Society},
  doi = {10.1103/PhysRevLett.133.236703}
}

@article{yuOddParityMagnetismDriven2025,
  title = {Odd-Parity Magnetism Driven by Antiferromagnetic Exchange},
  author = {Yu, Yue and Lyngby, Magnus B. and Shishidou, Tatsuya and Roig, Merc{\`e} and Kreisel, Andreas and Weinert, Michael and Andersen, Brian M. and Agterberg, Daniel F.},
  year = 2025,
  month = jul,
  journal = {Physical Review Letters},
  volume = {135},
  number = {4},
  pages = {046701},
  publisher = {American Physical Society},
  doi = {10.1103/zk69-k6b2}
}

@article{songElectricalSwitchingPwave2025,
  title = {Electrical Switching of a {$p$}-Wave Magnet},
  author = {Song, Qian and Stavri{\'c}, Srdjan and Barone, Paolo and Droghetti, Andrea and Antonenko, Daniil S. and Venderbos, J{\"o}rn W. F. and Occhialini, Connor A. and Ilyas, Batyr and Erge{\c c}en, Emre and Gedik, Nuh and Cheong, Sang-Wook and Fernandes, Rafael M. and Picozzi, Silvia and Comin, Riccardo},
  year = 2025,
  month = may,
  journal = {Nature},
  volume = {642},
  pages = {64--70},
  publisher = {Nature Publishing Group},
  doi = {10.1038/s41586-025-09034-7}
}

@article{chenUnconventionalMagnonsCollinear2025,
  title = {Unconventional Magnons in Collinear Magnets Dictated by Spin Space Groups},
  author = {Chen, Xiaobing and Liu, Yuntian and Liu, Pengfei and Yu, Yutong and Ren, Jun and Li, Jiayu and Zhang, Ao and Liu, Qihang},
  year = 2025,
  month = apr,
  journal = {Nature},
  volume = {640},
  number = {8058},
  pages = {349--354},
  publisher = {Nature Publishing Group},
  issn = {1476-4687},
  doi = {10.1038/s41586-025-08715-7},
  urldate = {2026-04-15},
  copyright = {2025 The Author(s)},
  langid = {english}
}

@article{cuiEfficientSpinSeebeck2023,
  title = {Efficient Spin Seebeck and Spin Nernst Effects of Magnons in Altermagnets},
  author = {Cui, Qirui and Zeng, Bowen and Cui, Ping and Yu, Tao and Yang, Hongxin},
  year = 2023,
  month = nov,
  journal = {Physical Review B},
  volume = {108},
  number = {18},
  pages = {L180401},
  publisher = {American Physical Society},
  doi = {10.1103/PhysRevB.108.L180401},
  urldate = {2026-07-01}
}

@article{liuChiralSplitMagnon2024a,
  title = {Chiral Split Magnon in Altermagnetic MnTe},
  author = {Liu, Zheyuan and Ozeki, Makoto and Asai, Shinichiro and Itoh, Shinichi and Masuda, Takatsugu},
  year = 2024,
  month = oct,
  journal = {Physical Review Letters},
  volume = {133},
  number = {15},
  pages = {156702},
  publisher = {American Physical Society},
  doi = {10.1103/PhysRevLett.133.156702},
  urldate = {2026-07-01}
}

@article{smejkalChiralMagnonsAltermagnetic2023,
  title = {Chiral Magnons in Altermagnetic {$\mathrm{RuO}_2$}},
  author = {{\v S}mejkal, Libor and Marmodoro, Alberto and Ahn, Kyo-Hoon and {Gonz{\'a}lez-Hern{\'a}ndez}, Rafael and Turek, Ilja and Mankovsky, Sergiy and Ebert, Hubert and D'Souza, Sunil W. and {\v S}ipr, Ond{\v r}ej and Sinova, Jairo and Jungwirth, Tom{\'a}{\v s}},
  year = 2023,
  month = dec,
  journal = {Physical Review Letters},
  volume = {131},
  number = {25},
  pages = {256703},
  publisher = {American Physical Society},
  doi = {10.1103/PhysRevLett.131.256703},
  urldate = {2025-07-29}
}

@article{wuMagnonSplittingTransport2025,
  title = {Magnon Splitting and Magnon Spin Transport in Altermagnets},
  author = {Wu, Kai and Dong, Jinyang and Zhu, Mingchao and Zheng, Fawei and Zhang, Jin-Hua},
  year = 2025,
  journal = {Chinese Physics Letters},
  volume = {42},
  number = {7},
  pages = {070702},
  doi = {10.1088/0256-307X/42/7/070702}
}

@article{jinInteractionDrivenAltermagnetic2026,
  title = {Interaction-Driven Altermagnetic Magnon Chiral Splitting},
  author = {Jin, Zhejunyu and Zeng, Zhaozhuo and Liu, Jie and Gong, Tianci and Su, Ying and Chang, Kai and Yan, Peng},
  year = 2026,
  month = feb,
  journal = {Physical Review Letters},
  volume = {136},
  number = {8},
  pages = {086703},
  publisher = {American Physical Society},
  doi = {10.1103/v867-h742}
}

@article{liuObservationSwitchableChiral2026,
  title = {Observation of {{Switchable Chiral Magnons}} in an {{Altermagnet}}},
  author = {Liu, Zheyuan and Kikuchi, Hodaka and Wei, Zijun and Asai, Shinichiro and Enderle, Mechthild and Hansen, Ursula B. and Garlea, Vasile O. and Le, Manh D. and Nilsen, Gøran J. and Zaliznyak, Igor A. and Masuda, Takatsugu},
  date = {2026-06-12},
  year = 2026,
  journal = {Physical Review Letters},
  shortjournal = {Phys. Rev. Lett.},
  volume = {136},
  number = {23},
  pages = {236705},
  doi = {10.1103/m8lc-f8gk}
}

@article{okaFloquetEngineeringQuantum2019,
  title = {Floquet Engineering of Quantum Materials},
  author = {Oka, Takashi and Kitamura, Sota},
  year = 2019,
  journal = {Annual Review of Condensed Matter Physics},
  volume = {10},
  pages = {387--408},
  doi = {10.1146/annurev-conmatphys-031218-013423}
}

@article{deLaTorreColloquiumNonthermal2021,
  title = {Colloquium: Nonthermal Pathways to Ultrafast Control in Quantum Materials},
  author = {{de la Torre}, A. and Kennes, D. M. and Claassen, M. and Gerber, S. and McIver, J. W. and Sentef, M. A.},
  year = 2021,
  month = dec,
  journal = {Reviews of Modern Physics},
  volume = {93},
  number = {4},
  pages = {041002},
  publisher = {American Physical Society},
  doi = {10.1103/RevModPhys.93.041002}
}

@article{huangLightInducedOddParityMagnetism2026,
  title = {Light-Induced Odd-Parity Magnetism in Conventional Antiferromagnetism},
  author = {Huang, Shengpu and Qin, Zheng and Zhan, Fangyang and Xu, Dong-Hui and Ma, Da-Shuai and Wang, Rui},
  year = 2026,
  month = mar,
  journal = {Physical Review Letters},
  volume = {136},
  number = {12},
  pages = {126703},
  publisher = {American Physical Society},
  doi = {10.1103/9346-9jpf},
  urldate = {2026-04-10}
}

@article{liFloquetSpinSplitting2026,
  title = {Floquet Spin Splitting and Spin Generation in Antiferromagnets},
  author = {Li, Bo and Shao, Ding-Fu and Kovalev, Alexey A.},
  year = 2026,
  month = apr,
  journal = {Physical Review Letters},
  volume = {136},
  number = {16},
  pages = {166701},
  issn = {0031-9007, 1079-7114},
  doi = {10.1103/xzm1-l6yf},
  urldate = {2026-05-08},
  langid = {english}
}

@article{zhuFloquetOddParityCollinear2026,
  title = {Floquet Odd-Parity Collinear Magnets},
  author = {Zhu, Tongshuai and Zhou, Di and Wang, Huaiqiang and Wei, Su-Huai and Ruan, Jiawei},
  year = 2026,
  month = mar,
  journal = {Physical Review Letters},
  volume = {136},
  number = {12},
  pages = {126704},
  publisher = {American Physical Society},
  doi = {10.1103/7ywb-ml2q},
  urldate = {2026-04-10}
}

@misc{zhangOddParityMagnons2026,
  title = {Odd-Parity Magnons},
  author = {Zhang, Pu and Xie, Sun-Bo and Yu, Junxi and Liu, Yichen and Liu, Cheng-Cheng},
  year = 2026,
  number = {arXiv:2605.31411},
  eprint = {2605.31411},
  publisher = {arXiv},
  doi = {10.48550/arXiv.2605.31411},
  urldate = {2026-07-01},
  copyright = {Creative Commons Attribution 4.0 International},
  langid = {english},
  archiveprefix = {arXiv}
}

@article{liuLightinducedOddparityAltermagnets2026,
  title = {Light-Induced Odd-Parity Altermagnets on Dimerized Lattices},
  author = {Liu, Dongling and Zhuang, Zheng-Yang and Zhu, Di and Wu, Zhigang and Yan, Zhongbo},
  year = 2026,
  month = feb,
  journal = {Physical Review B},
  volume = {113},
  number = {6},
  pages = {L060409},
  publisher = {American Physical Society},
  doi = {10.1103/wnqs-3djt},
  urldate = {2026-07-01}
}

@article{blochManybodyPhysicsUltracold2008,
  title = {Many-Body Physics with Ultracold Gases},
  author = {Bloch, Immanuel and Dalibard, Jean and Zwerger, Wilhelm},
  year = 2008,
  month = jul,
  journal = {Reviews of Modern Physics},
  volume = {80},
  number = {3},
  pages = {885--964},
  doi = {10.1103/RevModPhys.80.885}
}

@article{dasRealizingAltermagnetismFermiHubbard2024,
  title = {Realizing Altermagnetism in Fermi-Hubbard Models with Ultracold Atoms},
  author = {Das, Purnendu and Leeb, Valentin and Knolle, Johannes and Knap, Michael},
  year = 2024,
  month = jun,
  journal = {Physical Review Letters},
  volume = {132},
  number = {26},
  pages = {263402},
  doi = {10.1103/PhysRevLett.132.263402}
}

@article{eckardtColloquiumAtomicQuantum2017,
  title = {Colloquium: Atomic Quantum Gases in Periodically Driven Optical Lattices},
  author = {Eckardt, Andr{\'e}},
  year = 2017,
  month = mar,
  journal = {Reviews of Modern Physics},
  volume = {89},
  number = {1},
  pages = {011004},
  doi = {10.1103/RevModPhys.89.011004}
}

@article{gallegoMAGNDATADatabaseMagnetic2016,
  title = {MAGNDATA : Towards a Database of Magnetic Structures. I. The Commensurate Case},
  author = {Gallego, Samuel V. and {Perez-Mato}, J. Manuel and Elcoro, Luis and Tasci, Emre S. and Hanson, Robert M. and Momma, Koichi and Aroyo, Mois I. and Madariaga, Gotzon},
  year = 2016,
  month = oct,
  journal = {Journal of Applied Crystallography},
  volume = {49},
  number = {5},
  pages = {1750--1776},
  doi = {10.1107/S1600576716012863}
}

@article{goldmanPeriodicallyDrivenQuantum2014,
  title = {Periodically Driven Quantum Systems: Effective Hamiltonians and Engineered Gauge Fields},
  author = {Goldman, N. and Dalibard, J.},
  year = 2014,
  month = aug,
  journal = {Physical Review X},
  volume = {4},
  number = {3},
  pages = {031027},
  doi = {10.1103/PhysRevX.4.031027}
}

@article{jotzuExperimentalRealizationTopological2014,
  title = {Experimental Realization of the Topological Haldane Model with Ultracold Fermions},
  author = {Jotzu, Gregor and Messer, Michael and Desbuquois, R{\'e}mi and Lebrat, Martin and Uehlinger, Thomas and Greif, Daniel and Esslinger, Tilman},
  year = 2014,
  month = nov,
  journal = {Nature},
  volume = {515},
  number = {7526},
  pages = {237--240},
  doi = {10.1038/nature13915}
}

@article{miyakeRealizingHarperHamiltonian2013,
  title = {Realizing the Harper Hamiltonian with Laser-Assisted Tunneling in Optical Lattices},
  author = {Miyake, Hirokazu and Siviloglou, Georgios A. and Kennedy, Colin J. and Burton, William Cody and Ketterle, Wolfgang},
  year = 2013,
  month = oct,
  journal = {Physical Review Letters},
  volume = {111},
  number = {18},
  pages = {185302},
  doi = {10.1103/PhysRevLett.111.185302}
}

@misc{SM,
    note = {See Supplemental Material for detailed information on (i) the symmetry analysis establishing the complete classification of nonrelativistic odd-parity spin splitting for all 32 crystallographic point groups and the symmetry constraints on transport properties, (ii) the construction and symmetry analysis of electronic and magnonic models realizing the $p$-, $f_b$-, $f_p$-, $h$-, $k$-, and $m$-wave forms, and (iii) the Floquet protocol, material realizations, and computational details.}
  }

@article{kunnmannMagneticStructuresOrdered1968,
  title = {Magnetic Structures of the Ordered Trirutiles Cr2WO6, Cr2TeO6 and Fe2TeO6},
  author = {Kunnmann, W. and La Placa, S. and Corliss, L. M. and Hastings, J. M. and Banks, E.},
  year = 1968,
  month = aug,
  journal = {Journal of Physics and Chemistry of Solids},
  volume = {29},
  number = {8},
  pages = {1359--1364},
  issn = {0022-3697},
  doi = {10.1016/0022-3697(68)90187-X},
  urldate = {2026-07-13}
}

@article{mazetMagneticPropertiesMgFe6Ge62013,
  title = {Magnetic Properties of MgFe6Ge6},
  author = {Mazet, T. and Ban, V. and Sibille, R. and Capelli, S. and Malaman, B.},
  year = 2013,
  month = apr,
  journal = {Solid State Communications},
  volume = {159},
  pages = {79--83},
  issn = {0038-1098},
  doi = {10.1016/j.ssc.2013.01.027},
  urldate = {2026-07-13}
}

@article{searsAltermagneticDipolarSplitting2026,
  title = {Altermagnetic and Dipolar Splitting of Magnons in {$\mathrm{FeF}_2$}},
  author = {Sears, J. and Garlea, V. O. and Lederman, D. and Tranquada, J. M. and Zaliznyak, I. A.},
  year = 2026,
  month = jun,
  journal = {Physical Review Letters},
  volume = {136},
  number = {22},
  pages = {226701},
  publisher = {American Physical Society},
  doi = {10.1103/g6dt-rf8c},
  urldate = {2026-06-10}
}

@article{liuAntiferroaxialAltermagnetism2026a,
  title = {Antiferroaxial {{Altermagnetism}}},
  author = {Liu, Yichen and Liu, Cheng-Cheng},
  year = {2026},
  month = jun,
  journal = {Physical Review Letters},
  shortjournal = {Phys. Rev. Lett.},
  volume = {136},
  number = {25},
  pages = {256709},
  publisher = {American Physical Society},
  doi = {10.1103/21z4-c9p2}
}

@article{mcclartyLandauTheoryAltermagnetism2024,
  title = {Landau {{Theory}} of {{Altermagnetism}}},
  author = {McClarty, Paul A. and Rau, Jeffrey G.},
  year = {2024},
  month = apr,
  journal = {Physical Review Letters},
  shortjournal = {Phys. Rev. Lett.},
  volume = {132},
  number = {17},
  pages = {176702},
  publisher = {American Physical Society},
  doi = {10.1103/PhysRevLett.132.176702}
}

@article{schiffCollinearAltermagnetsTheir2025,
  title = {Collinear Altermagnets and Their {{Landau}} Theories},
  author = {Schiff, Hana and McClarty, Paul and Rau, Jeffrey G. and Romhányi, Judit},
  year = {2025},
  month = sep,
  journal = {Physical Review Research},
  shortjournal = {Phys. Rev. Research},
  volume = {7},
  number = {3},
  pages = {033301},
  issn = {2643-1564},
  doi = {10.1103/q44z-ynbr}
}

@article{yaoFirstPrinciplesCalculation2004,
  title = {First {{Principles Calculation}} of {{Anomalous Hall Conductivity}} in {{Ferromagnetic}} Bcc {{Fe}}},
  author = {Yao, Yugui and Kleinman, Leonard and MacDonald, A. H. and Sinova, Jairo and Jungwirth, T. and Wang, Ding-sheng and Wang, Enge and Niu, Qian},
  year = {2004},
  month = jan,
  journal = {Physical Review Letters},
  shortjournal = {Phys. Rev. Lett.},
  volume = {92},
  number = {3},
  pages = {037204},
  publisher = {American Physical Society},
  doi = {10.1103/PhysRevLett.92.037204}
}

@misc{xieGeneralTheoryChiral2026a,
  title = {A General Theory of Chiral Splitting of Magnons in Two-Dimensional Magnets},
  author = {Xie, Yu and Wang, Dinghui and Li, Chao and Shen, Xiaofan and Zhang, Junting},
  year = 2026,
  month = jan,
  number = {arXiv:2601.15031},
  eprint = {2601.15031},
  primaryclass = {cond-mat.mtrl-sci},
  publisher = {arXiv},
  doi = {10.48550/arXiv.2601.15031},
  urldate = {2026-07-20},
  archiveprefix = {arXiv}
}

@misc{elcoroSpinPointGroup2026,
  title = {Spin Point Group Symmetry and Classification of Non-Relativistic Spin Splitting in Non-Collinear Magnetic Structures: Identification of High-Order Spin Splitting Types ($\ell=5,7,$ and $9$)},
  author = {Elcoro, Luis and Etxebarria, Jesus and Perez-Mato, J. Manuel and Tasci, Emre S.},
  year = {2026},
  month = jun,
  number = {arXiv:2606.19254},
  eprint = {2606.19254},
  primaryclass = {cond-mat.mtrl-sci},
  publisher = {arXiv},
  doi = {10.48550/arXiv.2606.19254},
  url = {https://arxiv.org/abs/2606.19254},
  urldate = {2026-07-20},
  archiveprefix = {arXiv}
}
\appendix

\section*{End matter}
For a given equilibrium collinear spin point group, the N\'eel order transforms according to an irrep $\Gamma_N$ of the corresponding crystallographic point group. Circularly polarized light can be written as $\boldsymbol{\mathcal A}(t)=\operatorname{Re}[\boldsymbol{\mathcal A}_0e^{-i\omega t}]$, $\boldsymbol{\mathcal A}_0=\mathcal A_0(\hat{\mathbf x}+i\eta\hat{\mathbf y})$, where $\eta=\pm1$ 
specifies the helicity. The handedness of the drive is described by the time-reversal-odd axial vector $\boldsymbol\kappa=i\boldsymbol{\mathcal A}_0\times\boldsymbol{\mathcal A}_0^*$ which transforms according to the axial-vector representation $\Gamma_A$. The representation of the induced spin splitting therefore belongs to the product representation $\Gamma_\Delta\subseteq\Gamma_A\otimes\Gamma_N$. For each parent collinear spin point group, Table~\ref{table:floquet_coupling} gives $\Gamma_N$, the decomposition of $\Gamma_A$, and the resulting symmetry channels contained in $\Gamma_A\otimes\Gamma_N$. Table~\ref{table:odd_parity} then maps each allowed one-dimensional spin-splitting representation $\Gamma_\Delta$ onto its lowest-order odd-parity basis function.

For an odd spin splitting satisfying $\Delta E(-\bk)=-\Delta E(\bk)$, we define the partial-wave order $\ell$ as the degree of the lowest-order nonvanishing polynomial in the expansion of $\Delta E(\bk)$ around the $\Gamma$ point. The values $\ell=1,3,5,7,9$ correspond to the $p$-, $f$-, $h$-, $k$-, and $m$-wave classes, respectively. Among all crystallographic point groups, the maximal value is $\ell=9$, realized by the $A_{1u}$ representation of $m\bar{3}m$.

\begin{table*}[p]
\centering
\caption{\label{table:odd_parity}Lowest-order odd-parity spin-splitting basis functions for the one-dimensional irreducible representations $\Gamma_\Delta$ of all crystallographic point groups (PGs). The wave label is determined by the lowest odd polynomial degree $\ell=1,3,5,7$, or $9$. The point groups $1$, $3$, $4$, $4mm$, $23$, and $\bar{4}3m$ are omitted because none of their nontrivial one-dimensional irreps admits a symmetry-allowed odd-parity momentum-space basis at leading order.}
\footnotesize
\begin{ruledtabular}
\begin{tabular}{cll @{\hspace{0.7cm}} cll}
\textbf{PG} & \textbf{$\Gamma_\Delta$} & \textbf{Splitting [Basis function]}
& \textbf{PG} & \textbf{$\Gamma_\Delta$} & \textbf{Splitting [Basis function]} \\
\colrule
$\bar{1}$ & $A_u$ & $p$-wave $[x,y,z]$
& $\bar{3}$ & $A_u$ & $p$-wave $[z]$ \\
$2$ & $B$ & $p$-wave $[x,y]$
& $32$ & $A_2$ & $p$-wave $[z]$ \\
$m$ & $A''$ & $p$-wave $[x/y/z]$
& $3m$ & $A_2$ & $f$-wave $[x(x^2-3y^2)]$ \\
$2/m$ & $B_u$ & $p$-wave $[x,y]$
& $\bar{3}m$ & $A_{1u}$ & $f$-wave $[x(x^2-3y^2)]$ \\
 & $A_u$ & $p$-wave $[z]$
&  & $A_{2u}$ & $p$-wave $[z]$ \\
$222$ & $B_1$ & $p$-wave $[x]$
& $6$ & $B$ & $f$-wave $[x(x^2-3y^2)]$ \\
 & $B_2$ & $p$-wave $[y]$
& $\bar{6}$ & $A''$ & $p$-wave $[z]$ \\
 & $B_3$ & $p$-wave $[z]$
& $6/m$ & $A_u$ & $p$-wave $[z]$ \\
$mm2$ & $B_1$ & $p$-wave $[x]$
&  & $B_u$ & $f$-wave $[x(x^2-3y^2)]$ \\
 & $B_2$ & $p$-wave $[y]$
& $622$ & $A_2$ & $p$-wave $[z]$ \\
$mmm$ & $A_u$ & $f$-wave $[xyz]$
&  & $B_1$ & $f$-wave $[y(3x^2-y^2)]$ \\
 & $B_{1u}$ & $p$-wave $[x]$
&  & $B_2$ & $f$-wave $[x(x^2-3y^2)]$ \\
 & $B_{2u}$ & $p$-wave $[y]$
& $6mm$ & $B_1$ & $f$-wave $[y(3x^2-y^2)]$ \\
 & $B_{3u}$ & $p$-wave $[z]$
&  & $B_2$ & $f$-wave $[x(x^2-3y^2)]$ \\
$\bar{4}$ & $B$ & $p$-wave $[z]$
& $\bar{6}m2$ & $A_2''$ & $p$-wave $[z]$ \\
$4/m$ & $B_u$ & $f$-wave $[xyz,(x^2-y^2)z]$
&  & $A_2'$ & $f$-wave $[x(x^2-3y^2)]$ \\
 & $A_u$ & $p$-wave $[z]$
& $6/mmm$ & $A_{2u}$ & $p$-wave $[z]$ \\
$422$ & $A_2$ & $p$-wave $[z]$
&  & $A_{1u}$ & $k$-wave $[xyz(3x^2-y^2)(x^2-3y^2)]$ \\
$\bar{4}2m$ & $A_2$ & $f$-wave $[(x^2-y^2)z]$
&  & $B_{1u}$ & $f$-wave $[y(3x^2-y^2)]$ \\
 & $B_2$ & $p$-wave $[z]$
&  & $B_{2u}$ & $f$-wave $[x(x^2-3y^2)]$ \\
$4/mmm$ & $A_{2u}$ & $p$-wave $[z]$
& $m\bar{3}$ & $A_u$ & $f$-wave $[xyz]$ \\
 & $B_{1u}$ & $f$-wave $[(x^2-y^2)z]$
& $432$ & $A_2$ & $f$-wave $[xyz]$ \\
 & $B_{2u}$ & $f$-wave $[xyz]$
& $m\bar{3}m$ & $A_{2u}$ & $f$-wave $[xyz]$ \\
 & $A_{1u}$ & $h$-wave $[xyz(x^2-y^2)]$
&  & $A_{1u}$ & $m$-wave $[xyz(x^2-y^2)(y^2-z^2)(x^2-z^2)]$ \\
\end{tabular}
\end{ruledtabular}
\end{table*}

\begin{table*}[p]
\centering
\caption{\label{table:floquet_coupling}The coupling between a parent collinear N\'eel order and an axial vector introduced by an external field. The parent collinear spin point group (SPG) determines the collinear N\'eel order $\Gamma_N$. The full real-space axial-vector representation is denoted by $\Gamma_A$; its first one-dimensional component, when present, corresponds to the axial component along the principal axis. The product $\Gamma_A\otimes\Gamma_N$ gives the symmetry channels of the induced spin splitting.}
\scriptsize
\renewcommand{\arraystretch}{1.06}
\begin{ruledtabular}
\begin{tabular}{clll}
\textbf{PG} & \textbf{Axial vector $\Gamma_A$} & \textbf{Parent SPG ($\Gamma_N$)} & \textbf{$\Gamma_A\otimes\Gamma_N$} \\
\colrule
$\bar{1}$ & $3A_g$ & ${}^{\bar{1}}\bar{1}$ ($A_u$) & $3A_u$ \\
$2$ & $A\oplus2B$ & ${}^{\bar{1}}2$ ($B$) & $B\oplus2A$ \\
$m$ & $A'\oplus2A''$ & ${}^{\bar{1}}m$ ($A''$) & $A''\oplus2A'$ \\
$2/m$ & $A_g\oplus2B_g$ & ${}^{\bar{1}}2/m$ ($B_u$) & $B_u\oplus2A_u$ \\
 &  & $2/{}^{\bar{1}}m$ ($A_u$) & $A_u\oplus2B_u$ \\
$222$ & $B_1\oplus B_2\oplus B_3$ & $2^{\bar{1}}2^{\bar{1}}2$ ($B_1$) & $A\oplus B_2\oplus B_3$ \\
 &  & ${}^{\bar{1}}22^{\bar{1}}2$ ($B_2$) & $A\oplus B_1\oplus B_3$ \\
 &  & ${}^{\bar{1}}2^{\bar{1}}22$ ($B_3$) & $A\oplus B_1\oplus B_2$ \\
$mm2$ & $A_2\oplus B_1\oplus B_2$ & ${}^{\bar{1}}mm^{\bar{1}}2$ ($B_1$) & $A_1\oplus A_2\oplus B_2$ \\
 &  & $m^{\bar{1}}m^{\bar{1}}2$ ($B_2$) & $A_1\oplus A_2\oplus B_1$ \\
$mmm$ & $B_{1g}\oplus B_{2g}\oplus B_{3g}$ & ${}^{\bar{1}}m^{\bar{1}}m^{\bar{1}}m$ ($A_u$) & $B_{1u}\oplus B_{2u}\oplus B_{3u}$ \\
 &  & ${}^{\bar{1}}mmm$ ($B_{1u}$) & $A_u\oplus B_{2u}\oplus B_{3u}$ \\
 &  & $m^{\bar{1}}mm$ ($B_{2u}$) & $A_u\oplus B_{1u}\oplus B_{3u}$ \\
 &  & $mm^{\bar{1}}m$ ($B_{3u}$) & $A_u\oplus B_{1u}\oplus B_{2u}$ \\
$\bar{4}$ & $A\oplus E$ & ${}^{\bar{1}}\bar{4}$ ($B$) & $B\oplus E$ \\
$4/m$ & $A_g\oplus E_g$ & ${}^{\bar{1}}4/{}^{\bar{1}}m$ ($B_u$) & $B_u\oplus E_u$ \\
 &  & $4/{}^{\bar{1}}m$ ($A_u$) & $A_u\oplus E_u$ \\
$422$ & $A_2\oplus E$ & $4^{\bar{1}}2^{\bar{1}}2$ ($A_2$) & $A_1\oplus E$ \\
$\bar{4}2m$ & $A_2\oplus E$ & $\bar{4}^{\bar{1}}2^{\bar{1}}m$ ($A_2$) & $A_1\oplus E$ \\
 &  & ${}^{\bar{1}}\bar{4}^{\bar{1}}2m$ ($B_2$) & $B_1\oplus E$ \\
$4/mmm$ & $A_{2g}\oplus E_g$ & $4/{}^{\bar{1}}mmm$ ($A_{2u}$) & $A_{1u}\oplus E_u$ \\
 &  & ${}^{\bar{1}}4/{}^{\bar{1}}mm^{\bar{1}}m$ ($B_{1u}$) & $B_{2u}\oplus E_u$ \\
 &  & ${}^{\bar{1}}4/{}^{\bar{1}}m^{\bar{1}}mm$ ($B_{2u}$) & $B_{1u}\oplus E_u$ \\
 &  & $4/{}^{\bar{1}}m^{\bar{1}}m^{\bar{1}}m$ ($A_{1u}$) & $A_{2u}\oplus E_u$ \\
$\bar{3}$ & $A_g\oplus E_g$ & ${}^{\bar{1}}\bar{3}$ ($A_u$) & $A_u\oplus E_u$ \\
$32$ & $A_2\oplus E$ & $3^{\bar{1}}2$ ($A_2$) & $A_1\oplus E$ \\
$3m$ & $A_2\oplus E$ & $3^{\bar{1}}2$ ($A_2$) & $A_1\oplus E$ \\
$\bar{3}m$ & $A_{2g}\oplus E_g$ & ${}^{\bar{1}}\bar{3}^{\bar{1}}m$ ($A_{1u}$) & $A_{2u}\oplus E_u$ \\
 &  & ${}^{\bar{1}}\bar{3}m$ ($A_{2u}$) & $A_{1u}\oplus E_u$ \\
$6$ & $A\oplus E_1$ & ${}^{\bar{1}}6$ ($B$) & $B\oplus E_2$ \\
$\bar{6}$ & $A'\oplus E''$ & ${}^{\bar{1}}\bar{6}$ ($A''$) & $A''\oplus E'$ \\
$6/m$ & $A_g\oplus E_{1g}$ & $6/{}^{\bar{1}}m$ ($A_u$) & $A_u\oplus E_{1u}$ \\
 &  & ${}^{\bar{1}}6/m$ ($B_u$) & $B_u\oplus E_{2u}$ \\
$622$ & $A_2\oplus E_1$ & $6^{\bar{1}}2^{\bar{1}}2$ ($A_2$) & $A_1\oplus E_1$ \\
 &  & ${}^{\bar{1}}62^{\bar{1}}2$ ($B_1$) & $B_2\oplus E_2$ \\
 &  & ${}^{\bar{1}}6^{\bar{1}}22$ ($B_2$) & $B_1\oplus E_2$ \\
$6mm$ & $A_2\oplus E_1$ & ${}^{\bar{1}}6m^{\bar{1}}m$ ($B_1$) & $B_2\oplus E_2$ \\
 &  & ${}^{\bar{1}}6^{\bar{1}}mm$ ($B_2$) & $B_1\oplus E_2$ \\
$\bar{6}m2$ & $A_2'\oplus E''$ & ${}^{\bar{1}}\bar{6}m^{\bar{1}}2$ ($A_2''$) & $A_1''\oplus E'$ \\
 &  & $\bar{6}^{\bar{1}}m^{\bar{1}}2$ ($A_2'$) & $A_1'\oplus E''$ \\
$6/mmm$ & $A_{2g}\oplus E_{1g}$ & $6/{}^{\bar{1}}mmm$ ($A_{2u}$) & $A_{1u}\oplus E_{1u}$ \\
 &  & $6/{}^{\bar{1}}m^{\bar{1}}m^{\bar{1}}m$ ($A_{1u}$) & $A_{2u}\oplus E_{1u}$ \\
 &  & ${}^{\bar{1}}6/mm^{\bar{1}}m$ ($B_{1u}$) & $B_{2u}\oplus E_{2u}$ \\
 &  & ${}^{\bar{1}}6/m^{\bar{1}}mm$ ($B_{2u}$) & $B_{1u}\oplus E_{2u}$ \\
$m\bar{3}$ & $T_g$ & ${}^{\bar{1}}m^{\bar{1}}\bar{3}$ ($A_u$) & $T_u$ \\
$432$ & $T_1$ & ${}^{\bar{1}}43^{\bar{1}}2$ ($A_2$) & $T_2$ \\
$m\bar{3}m$ & $T_{1g}$ & ${}^{\bar{1}}4/{}^{\bar{1}}m^{\bar{1}}\bar{3}^{\bar{1}}2/m$ ($A_{2u}$) & $T_{2u}$ \\
 &  & $4/{}^{\bar{1}}m^{\bar{1}}\bar{3}2/{}^{\bar{1}}m$ ($A_{1u}$) & $T_{1u}$ \\
\end{tabular}
\end{ruledtabular}
\end{table*}

\end{document}